\documentclass[useAMS,usenatbib,onecolumn,usegraphicx]{mn2e}

\newcommand{\gsim}{\, \raisebox{-0.8ex}{$\stackrel{\textstyle >}{\sim}$ }}

\newcommand{\beq}{\begin{equation}}
\newcommand{\eeq}{\end{equation}}
\newcommand{\beqar}{\begin{eqnarray}}
\newcommand{\eeqar}{\end{eqnarray}}

\title[Massive Hybrid Quark Stars]
{Massive Hybrid Quark Stars with Strong Magnetic Field}
\author[H. Sotani \& T. Tatsumi]
{Hajime Sotani$^1$ \thanks{E-mail:sotani@yukawa.kyoto-u.ac.jp},
and
Toshitaka Tatsumi$^2$
\\
$^1$Division of Theoretical Astronomy, National Astronomical Observatory of Japan, 2-21-1 Osawa, Mitaka, Tokyo 181-8588, Japan\\
$^2$Department of Physics, Kyoto University, Kyoto 606-8502, Japan}

\begin{document}
\maketitle
\label{firstpage}

\begin{abstract}
We estimate the critical magnetic field strength at which the lowest or second Landau levels play an important role in the quark phase inside the hybrid stars, and show that the magnetic field should be in the order of $10^{19}$ Gauss at the nuclear density $\sim0.16$ fm$^{-3}$. We also find that the pressure of quark matter settling only in the lowest Landau level can be expressed as a function of the energy density independently of the magnetic field strength, which corresponds to the causality limit of a stiff equation of state. Adiabatic index of quark matter well exceeds 4/3 in the core, and we find a possibility to construct massive hybrid star models occupied in large part by quark matter, whose maximum mass becomes larger than two solar mass.
\end{abstract}


\section{Introduction}
\label{sec:I}

Neutron stars must be one of the most suitable environments to examine the physics under extreme conditions. The density inside the star can be  much more than the nuclear saturation density, $\varepsilon_0=2.7\times 10^{14}$ g/cm$^3$, and its gravitational binding energy becomes too large, such as $M/R\simeq 0.2$ with the stellar mass ($M$) and radius ($R$) \citep{NS}. Via the observations of the phenomena associated with the compact objects, one might reveal not only the equation of state (EOS) and nuclear properties (e.g., \citet{AK1996,STM2001,SH2003,SKH2004,PA2011,DGKK2013,SNIO2012,SNIO2013a,SNIO2013b}), but also the gravitational theory itself (e.g., \citet{SK2004,SK2005,S2009a,S2009b,YYT2012,S2014a,S2014b,S2014c}). In addition, the observations of pulsars tell us that the neutron stars generally have a strong magnetic field, such as $\sim 10^{12}-10^{13}$ Gauss \citep{Pulsar}. Furthermore, the existence of another class of neutron stars, the so-called magnetars, is also suggested observationally \citep{DT1992,TD1993,TD1996}. The surface magnetic field of magnetars can reach as large as $10^{14}-10^{15}$ Gauss, which is determined through the measurements of the rotational period and spin down rate of soft gamma repeaters and anomalous X-ray pulsars \citep{K1998,H1999,M1999}. According to the population statistics of soft gamma repeaters, even $\sim 10 \%$ of the neutron stars produced via supernovae are expected to be magnetars \citep{K1998}.

The origin of such a strong magnetic field of neutron star is still uncertain. Maybe, a simple model is caused by fossil magnetic field of progenitor star. That is, a small magnetic field in progenitor star would be amplified during the gravitational collapse with conserving the magnetic flux, which produces the strong magnetic field of neutron star \citep{C1992}. Unfortunately, this hypothesis could be difficult to be accepted for magnetars, because the stellar radius with the canonical mass of $M\approx 1.4M_\odot$ should be less than the Schwarzshild radius defined by $R_{\rm Sch} = 2GM/c^2$ to produce a strong surface magnetic field such as $\sim 10^{15}$ Gauss \citep{tatsumi00}. Another possibility of the generation mechanism is due to the magnetohydrodynamic dynamo mechanism, i.e., the rapidly rotation of protoneutron star with the rotational period smaller than 3 ms may amplify a seed magnetic field up to $\sim 10^{15}$ Gauss \citep{DT1992,TD1993}. This scenario is also unfavorable from the observations of supernova remnants associated with the magnetar candidates \citep{VK2006}. Additionally, the possibility of the ferromagnetism of quark liquid inside the neutron star is suggested as an origin of a strong magnetic field \citep{tatsumi00}. Conceivably, the hint to solve the open question about the origin of strong magnetic field of neutron star, might lay behind the magnetized properties in the core region.

Furthermore, the exact structure of neutron stars is still unclear, because the EOS of neutron star matter, especially for high density region, is not fixed yet. Even so, it is believed that, under the surface ocean composed of molten metal, the neutron-rich nuclei form a lattice structure through the Coulomb interaction, which is usually called a crust region \citep{LP2004}. As the energy density increases up to $\sim \varepsilon_0$, the nuclei in the crust region would melt into uniform matter. This region corresponds to the core of neutron stars. Moreover, non-hadronic matter, such as quarks, might appear in deeper part of core region, depending on the theoretical model of neutron star matter \citep{NS}. In particular, the neutron star having quark matter is referred to as the hybrid star. Inside the hybrid star, one has only to consider three flavor quark matter composed of $u$, $d$, and $s$ quarks, because the more massive quarks can not be produced under the typical energy density of neutron stars. Since quark matter makes EOS soft, the mass of hybrid star is usually expected to be small. In practice, most of the hybrid stars proposed up to now might be difficult to reach the observed maximum mass, which is about two solar mass \citep{D2010,A2013}. Or, quark matter occupies very tiny region, even if the massive hybrid stars could be constructed.

Along with the origin of magnetic field of neutron star, the strength of magnetic field inside the star is also uncertain. 
The dipole magnetic field must dominate outside the star, while the magnetic configuration inside the star may be more complex due to the magnetic instability. According to the virial theorem, the maximum magnetic field for the neutron star with $R\simeq 10$ km and $M\simeq 1.4M_\odot$ could be in the order of $\sim 10^{18}$ Gauss \citep{LS1991}. On the other hand, the maximum magnetic field in the quark phase might reach $\sim 10^{20}$ Gauss, adopting the condition that the magnetic energy density should not exceed the energy density of the self-bound quark matter \citep{Ferrer2010}. If the magnetic field inside the star would be so strong, one has to take into account the effect of magnetic field on the neutron star matter, where the quantum effects such as the Landau levels may also play an important role to determine the stellar configuration. So far, there are a few literatures about the macroscopic structures of hybrid stars under the magnetic field \citep{Rabhi2009,CCM2014}. Considering the effects of the magnetic field in hadronic and quark phases, we find two important differences. One is the population of the charged particles: neutrons dominate over charged particles in the hadronic phase, while all the quarks have net electric charge. The other one is the difference of the mass of particles: baryons have much larger mass than quarks. These two features suggest that the effect of the magnetic field should be more important in quark matter. In this paper, we will consider the Landau level in the quark phase of hybrid star. We derive a critical magnetic field $B_c$ for given density, above which only the lowest Landau level is occupied, and find the EOS becomes stiffest as a causality limit for $B>B_c$, independent of the magnetic field. In addition, we especially examine how massive the hybrid star can become with the effect of strong magnetic field. We demonstrate that the EOS becomes {\it stiff} as a result of the hadron-quark phase transition in the strong magnetic field.
Then, we find that the hybrid star whose mass is more than two solar mass, can have a very large quark core, i.e., ``{\it hybrid quark star}". Unless otherwise noted, we adopt the geometric unit of $c=G=1$, where $c$ and $G$ denote the speed of light and the gravitational constant, respectively.

\section{Effects of Magnetic fields}
\label{sec:II}

Assuming the uniform magnetic field along the $z$ axis, the $n$-th energy level of quark with flavor $f$ in the strongly magnetic field is given by 
\begin{equation}
  E_n^f = \sqrt{c^4m_f^2 + c^2p_z^2 +  \hbar c |e_fB|[2n+1+{\rm sgn}(e_f B)s]}, \label{eq:En}
\end{equation}
where $m_f$, $\hbar$, $B$, and $s$ denote the particle mass, the Plank constant, the magnetic field strength, and the degree of freedom of spin, respectively, while $e_f=(2e/3,-e/3,-e/3)$ with the electron charge $e$. As shown later, the lowest Landau level (LLL) plays a primary role in our discussion, where quark matter settles only in LLL, the quark number density $n_{f}$ is given by
\begin{equation}
  n_{f} = \frac{3|e_fB|}{2\pi^2 \hbar^2 c}p_{f\rm F}  \label{eq:nb}
\end{equation}
where $p_{f\rm F}$ denotes the Fermi momentum of quark. Since the typical values of the number density and magnetic field are $n_f\simeq 3n_0$ with the saturation density $n_0=0.16$ fm$^{-3}$ and $B\simeq 10^{18}$ Gauss in this paper, the Fermi momentum of quark can be estimated as $p_{f\rm F}\simeq 4\times |e/e_f|$ GeV, which is much less than $m_f$. Then, the mass term in Eq. (\ref{eq:En}) does not significantly contribute to the energy level, $E_n$. In this case, the energy density of quark $\varepsilon_f$ is 
\begin{equation}
   \varepsilon_f = \frac{3|e_fB|}{4\pi^2\hbar^2}p_{f\rm F}^2,  \label{eq:epsilon}
\end{equation}
and quark matter becomes flavor symmetric, i.e., $n_u\simeq n_d\simeq n_s$, which leads to $2p_{u\rm F}\simeq p_{d\rm F}\simeq p_{s\rm F}$. We remark that the baryon number density $n_{\rm b}$ is given by $n_{\rm b}=(n_u+n_d+n_s)/3$. Then, the total energy density $\varepsilon$ is defined by $\varepsilon = \varepsilon_u+\varepsilon_d+\varepsilon_s + {\cal B}$ within the MIT bag model, 
where ${\cal B}$ denotes the bag constant. From Eqs. (\ref{eq:nb}) and (\ref{eq:epsilon}), $\varepsilon$ is 
\begin{equation}
   \varepsilon = \frac{5\pi^2 \hbar^2 c^2}{2eB}n_{\rm b}^2 + {\cal B},
\end{equation}
while the pressure, $P$, is calculated by
\begin{equation}
  P=n_{\rm b}^2\frac{\partial (\varepsilon/n_{\rm b})}{\partial n_{\rm b}}.
\end{equation}
As a result, one can express $P$ as a function of $n_{\rm b}$, such as
\begin{eqnarray}
  P &=& \frac{5\pi^2\hbar^2c^2}{2eB}n_{\rm b}^2 - {\cal B}  \nonumber \\
    &=& 1.31 \times 10^{35}\ {\rm erg/cm}^3 \times \left(\frac{n_{\rm b}}{n_0}\right)^2 {B_{19}}^{-1} - {\cal B} \nonumber \\
    &=& 82.0\ {\rm MeV/fm}^3 \times \left(\frac{n_{\rm b}}{n_0}\right)^2 {B_{19}}^{-1} - {\cal B}   \label{eq:P}
\end{eqnarray}
where $B_{19}$ is defined by $B_{19}=B/(10^{19}$ Gauss). Although the pressure expressed by Eq. (\ref{eq:P}) depends on $B$ as well as the baryon number density, we can find that the relation between $P$ and $\varepsilon$ becomes that $P=\varepsilon - 2{\cal B}$, which is independently of the magnetic field strength. From this relationship, the adiabatic speed of sound, $c_{\rm s}=(dP/d\varepsilon)^{1/2}$, becomes equivalent to the speed of light $c$. Thus, this is corresponding to the limiting case of a stiff EOS, because $c_{\rm s}$ can not exceed $c$ to keep the causality. Note that the EOS of free quark matter gives $c_{\rm s}=c/\sqrt{3}$. We also remark that our EOS would reduce to the same as the EOS of free quark matter in the limit of low magnetic field, where quarks occupy all the Landau levels.

\begin{figure*}
\begin{center}
\begin{tabular}{cc}
\includegraphics[scale=0.4]{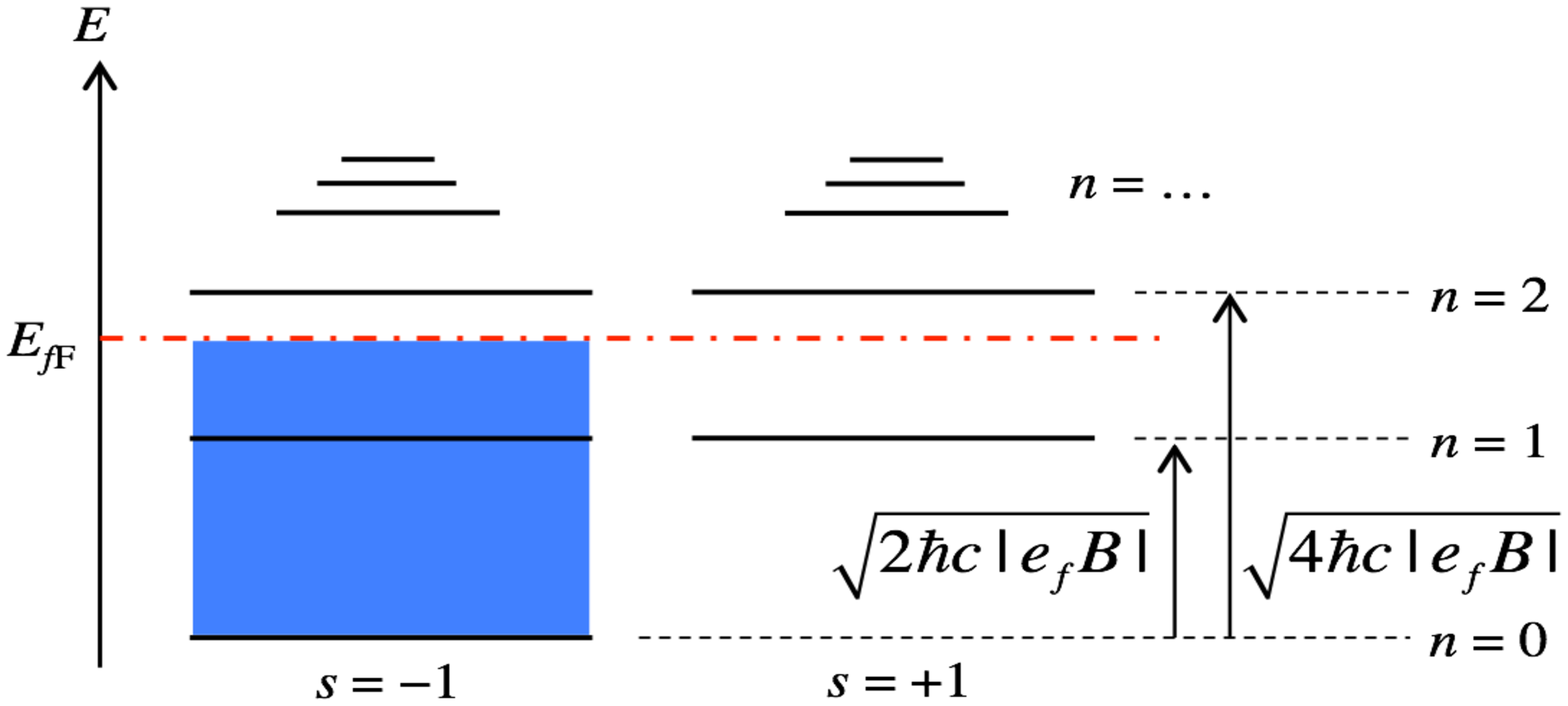} &
\includegraphics[scale=0.4]{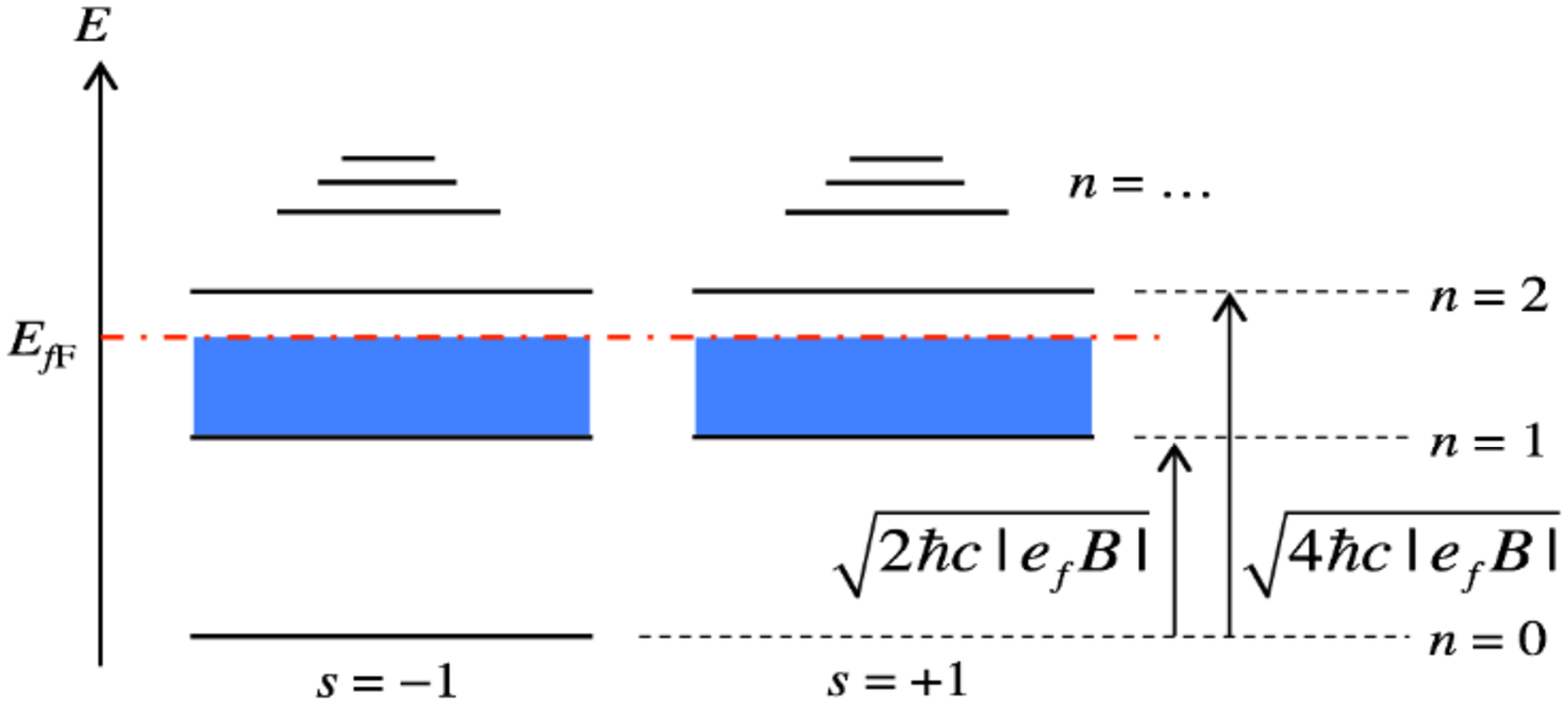}
\end{tabular}
\end{center}
\caption{
Image of the energy level for magnetized quark matter. In the case that the Fermi energy of quark matter with flavor $f$ is larger than $E_1^f$ and less than $E_2^f$ ($E_1^f<E_{\rm F}^f< E_2^f$), quark matter exists in the state either with the momentum $p_{f\rm F}^{(1)}$ above the lowest energy level $E_0^f$ (left panel) or with $p_{f\rm F}^{(2)}$ above the first excited energy $E_1^f$ (right panel).
}
\label{fig:LLL}
\end{figure*}

The condition that quark matter can settle only in LLL is that $E_{f\rm F} < \sqrt{2\hbar c |e_fB|}$, where $E_{f\rm F}$ denotes the Fermi energy of quark. With the relation of $E_{f\rm F}=cp_{f\rm F}$ in LLL, one can derive the critical strength of the magnetic field so that quarks only occupy LLL for each flavor. We then find that the critical strength for $u$ quarks is weaker than that for $d$ and $s$ quarks, due to the difference of electric charge $e_f$. Thus, the condition for all quarks only occupy LLL is given by that the magnetic field should be stronger than the critical strength for $d$ and $s$ quarks, i.e., one can show that the magnetic field strength should be larger than the critical strength $B_c$ given by
\begin{eqnarray}
   B_c &=& \frac{\sqrt[3]{6\pi^4}\hbar c}{e}n_{\rm b}^{2/3} \nonumber \\
           &=& 1.62\times 10^{19} \times\left(\frac{n_{\rm b}}{n_0}\right)^{2/3} \ \rm{Gauss}.    \label{eq:Bc1}
\end{eqnarray}
It may be interesting to see that this critical strength is not changed even in the non-relativistic limit. Thus the critical strength can be drawn as a universal function of density (Fig.~\ref{fig:LLL}). If the magnetic field exceeds $B_c$ at some density region in quark matter, EOS is given by the causal limit there. Note that EOS becomes invariant for the increase of the magnetic field once $B>B_c$ holds.

\begin{figure}
\begin{center}
\includegraphics[scale=0.5]{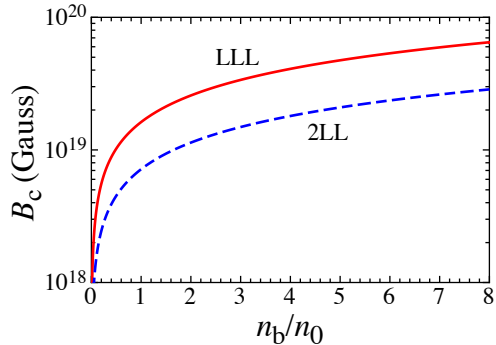} 
\end{center}
\caption{
The critical magnetic field strengths so that quark matter settles only in LLL and up to the 2nd Landau level (2LL) are shown as a function of the baryon number density $n_{\rm b}$, where the solid and broken lines correspond to the results for LLL and 2LL, respectively.
}
\label{fig:Bc}
\end{figure}

The situation that quark matter can settle up to the 2nd Landau level is more complicated. That is, as shown in Fig. \ref{fig:LLL}, the magnetized quark matter with Fermi energy $E_{f\rm F}$ can exist in the state either with the momentum $p_{f\rm F}^{(1)}$ above the lowest energy level $E_0^f$ or with $p_{f\rm F}^{(2)}$ above the first excited energy $E_1^f$, where $E_{f\rm F}$ is given by
\begin{equation}
   E_{f\rm F} = cp_{f\rm F}^{(1)} = \left(c^2 {p_{f\rm F}^{(2)}}^2 + 2\hbar c |e_fB|\right)^{1/2}.    \label{eq:EF2}
\end{equation}
In this situation, the quark number density and the energy density of quark can be written as
\begin{eqnarray}
  n_{f} &=& \frac{3|e_fB|}{2\pi^2\hbar^2c}\left[p_{f\rm F}^{(1)} + 2p_{f\rm F}^{(2)}\right], \label{eq:nb2} \\
  \varepsilon_f &=& \frac{3|e_fB|}{4\pi^2\hbar^2}\left[{p_{f\rm F}^{(1)}}^2 + 2p_{f\rm F}^{(1)}p_{f\rm F}^{(2)} 
     + \frac{4\hbar |e_f B|}{c}\ln\left|p_{f\rm F}^{(1)} + p_{f\rm F}^{(2)}\right| - \frac{2\hbar |e_fB|}{c}\ln\left(\frac{2\hbar |e_fB|}{c}\right)\right].   \label{eq:e2}
\end{eqnarray}
One expects that, due to the increase of the degree of freedom, the EOS in this situation could become somewhat softer than the result obtained when quark matter exists only in LLL, i.e., $P=\varepsilon - 2{\cal B}$. In practice, one may be able to derive the explicit relation between $P$ and $\varepsilon$ with the above Eqs. (\ref{eq:EF2}) -- (\ref{eq:e2}), but here we avoid to derive a complicated expression, because we focus on the stiffest case of EOS with the effects of the Landau levels in this paper. Instead of the explicit expression of EOS, we only point out how the critical value of magnetic field strength decreases in this case. Since the condition that quark matter settles up to the 2nd Landau level should be that $E_{f\rm F} < \sqrt{4\hbar c |e_fB|}$, one can show that the critical magnetic field is given by
\begin{eqnarray}
  B_c &=& \left[\frac{3\pi^4}{(1+\sqrt{2})^2}\right]^{1/3}\frac{\hbar c}{e}n_{\rm b}^{2/3} \nonumber \\
          &=& 7.15\times 10^{18} \times\left(\frac{n_{\rm b}}{n_0}\right)^{2/3} \ \rm{Gauss}.   \label{eq:Bc2}
\end{eqnarray}
The critical field strengths $B_c$ so that quark matter settles only in the LLL and up to the 2nd Landau level given by Eqs. (\ref{eq:Bc1}) and (\ref{eq:Bc2}), are shown as a function of the baryon number density in Fig. \ref{fig:Bc}. Considering that the baryon number density at the center of neutron star would be the order of $\sim (5-7)n_0$ fm$^{-3}$, the effects of the Landau level become considerably important only when the magnetic field strength is larger than $\sim 10^{19}$ Gauss. On the other hand, although the distribution and strength of magnetic field inside the star are still unclear, the magnetic field may reach such a large strength in the neutron star core composed of  quark matter. We also remark that in any case the simple dipole magnetic distribution inside the star can not realize enough large strength that the effects of the Landau levels play important roles in the neutron star core (see for the detailed discussion in Appendix \ref{sec:appendix_1}).

\section{Hybrid Star Models}
\label{sec:III}

Now, we consider how the properties of hybrid stars could be changed due to the effects of the Landau levels, when the magnetic field would be enough strong in the stellar core. In particular, we focus on the case that quark matter settles only in LLL, because such case realizes the stiffest EOS as mentioned in the previous section. The equilibrium configuration of magnetized neutron star is generally deformed due to the nonradial magnetic pressure. 
But, as a first step, we simply neglect the effect of magnetic pressure on the stellar configuration in this paper,
i.e., the stellar configuration becomes spherically symmetric, because the structure of magnetic field inside the star is still unknown and the stellar deformation strongly depends on the magnetic geometry \citep{BBGN1995}.
So, the stellar models considered in this paper are determined by solving the so-called Tolman-Oppenheimer-Volkof equations together with the relationship between the total energy density and pressure, i.e., the EOS.

Here, we should remark the magnetohydrodynamic issues associated with such a strong magnetic field advocated in this paper. \cite{CC1968} considered the quantum theory of a relativistic electron gas in the magnetic field, and pointed out that the kinetic parallel and perpendicular pressures with respect to the  direction of the magnetic field could be anisotropic. However, \cite{BH1982} have shown that the pressure is always isotropic due to the magnetization currents in compressed plasma, through the calculations of the magnetic susceptibility for degenerate free electrons in the crust of a neutron star. Recently, \cite{PY2012} have also shown that the hydrostatic equilibrium of a volume element in a magnetized star does not depend on the direction of the magnetic field, i.e., the pressure is isotropic. On the other hand, \cite{HHRS2010} discussed the mechanical stability of  a system formed by quarks confined to their lowest Landau level, and the possibility that the transverse pressure tends to vanish under such a strong magnetic field, which may induce a gravitational collapse. Anyway, these issues might be more important under the strong magnetic field considered in this paper, which would be discussed elsewhere.

\begin{figure}
\begin{center}
\includegraphics[scale=0.5]{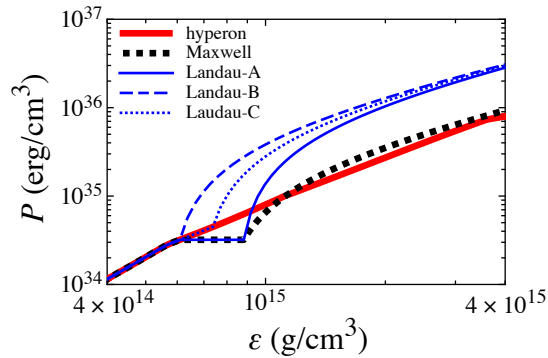} 
\end{center}
\caption{
Relationship between the total energy density ($\varepsilon$) and pressure ($P$) for EOSs adopted in this paper (see text for details). 
}
\label{fig:EOS}
\end{figure}

\begin{figure}
\begin{center}
\includegraphics[scale=0.5]{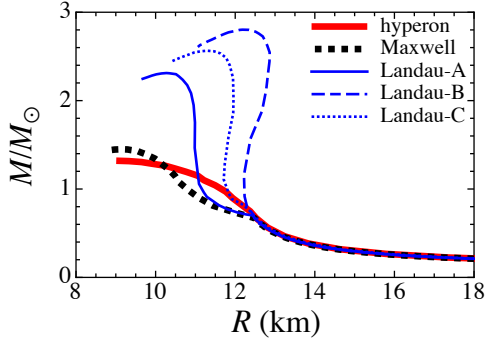} 
\end{center}
\caption{
Mass-radius relations of neutron stars constructed with several EOSs shown in Fig. \ref{fig:EOS}.
}
\label{fig:MR}
\end{figure}

Before considering the stellar models with the effects of the Landau level, for reference, we construct the stellar model with the same EOSs as in \citet{TSHT2007,Yasutake2009,SYMT2011}. That is, for hadronic matter, we adopt the EOS  based on the non-relativistic Brueckener-Hartree-Fock approach with $\Sigma^-$ and $\Lambda$ hyperons \citep{Baldo1998}, which is referred to as ``hyperon EOS" in this paper. For the quark phase, we adopt the sophisticated MIT bag model suggested in \citet{yasutake09a,chen09}, which are composed of the massless $u$ and $d$ quarks and $s$ quark with the current mass of $m_{\rm s}=150$ MeV, where the bag constant is set to be 100 MeV fm$^{-3}$. Then, the quark phase is connected to the hadronic matter with a Maxwell construction. The resultant EOS is referred to as ``Maxwell EOS," where the energy density becomes discontinuous between $5.93 \times 10^{14}$ and $8.82 \times 10^{14}$ g/cm$^3$. We show the pressure as a function of the total energy density for hyperon and Maxwell EOSs in Fig. \ref{fig:EOS}, while the corresponding mass-radius relations of the constructed neutron stars in Fig. \ref{fig:MR}, where the thick-solid and thick-dotted lines denote the results with hyperon and Maxwell EOSs, respectively. From Fig. \ref{fig:EOS}, one can see that the quark phase becomes stiffer than the hadronic matter in the high density region. As a result, the stellar models with quark core can be more massive than those without quark core, as in Fig. \ref{fig:MR}. However, the maximum mass of the stellar model with quark core is still too small to explain the observed maximum mass, i.e., two-solar mass, which is a big problem for considering hybrid stars.

Next, we consider the hybrid star models with the effects of LLL. For this purpose, we consider that the quark phase in Maxwell EOS would be modified due to the existence of strong magnetic field. In the high density region, such a modified EOS should be expressed as $P=\varepsilon - 2{\cal B}$, as derived in the previous section. Meanwhile, it is not clear how the EOS for quark matter would be connected with the hadronic matter at the moderate density. So, in this paper, we adopt three possible cases to construct the modified EOS. That is, 
the EOS for quark matter is connected with the hadronic matter (A) at the upper limit of the density discontinuity in Maxwell EOS, i.e., $\varepsilon_u = 8.82 \times 10^{14}$ g/cm$^3$, (B) at the lower limit of the density discontinuity in Maxwell EOS, i.e., $\varepsilon_l = 5.93 \times 10^{14}$ g/cm$^3$, and (C) at the density defined as $\varepsilon_m=(\varepsilon_u + \varepsilon_l)/2$, as shown in Fig. \ref{fig:EOS}. Probably, almost all possibilities of how to connect the EOS for quark matter modified by the strong magnetic field with the hadronic EOS must be covered by the above cases from (A) to (C), although the value of ${\cal B}$ might be different from the value of bag constant in Maxwell EOS because of the existence of magnetic field.
In fact, such a connection between quark and hadronic matter has effectively shifted the value of the bag constant into ${\cal B}= 237.3$ MeV fm$^{-3}$ for the case (A), ${\cal B}= 160.9$ MeV fm$^{-3}$ for the case (B), and ${\cal B}= 192.8$ MeV fm$^{-3}$ for the case (C).
Anyway, we believe that the three cases are sufficient to see the qualitative behavior of the hybrid star models with the effects of the Landau level. Hereafter, we refer to these modified EOSs as ``Landau-A," ``Landau-B," and ``Landau-C" EOSs.

Fig. \ref{fig:MR} shows the mass-radius relations of the hybrid stars constructed with the Landau-A (solid line), Landau-B (broken line), and Landau-C EOSs (dotted line). From this figure, we find that the maximum mass of the hybrid star becomes larger for the stellar model with EOS connected to the hadronic matter at the lower energy density, i.e., the maximum masses are $2.80M_\odot$ for the Landau-B EOS, $2.56M_\odot$ for the Landau-C EOS, and $2.31M_\odot$ for the Landau-A EOS. Since the Landau-A EOS is the softest among the three EOSs due to the existence of a large discontinuity in energy density, the maximum mass for the Landau-A EOS becomes smaller than the other cases. Nevertheless, one can obviously see that in any cases with the effect of LLL, the maximum masses become much larger than that constructed with the Maxwell EOS. As a result, these models can avoid the observational problem, i.e., the expected maximum masses are larger than the two-solar mass. Up to now, the explanation of the two-solar mass with the hybrid stars is difficult, but we are successful to show the possibility to construct the massive stellar models by considering the effect of the strong magnetic field. Here, we remark that the strong magnetic field is necessary at least in the core region to construct such a massive hybrid star, but the magnetic field in the crust region and at the stellar surface does not take so large (maybe at most $B\sim 10^{16}$ Gauss). It should be worthwhile to discuss the stability of hybrid stars near the maximum mass. Since quark matter is well approximated by the free quark-gas in the core region due to the asymptotic freedom of QCD, we can discuss the stability in a rather model-independent way. It is well-known that the adiabatic index $\gamma$ of the free quark gas is $4/3$, and cannot satisfy the stability criterion for compact stars, $\gamma>4/3+\kappa M/R$ with $\kappa\sim O(1)$ \citep{shapiro-teukolsky}. However, we find that the adiabatic index of the EOS of quark matter becomes almost $2$ from Eq.~(\ref{eq:P}) in the presence of the magnetic field, which value satisfies the criterion.

\begin{figure}
\begin{center}
\includegraphics[scale=0.5]{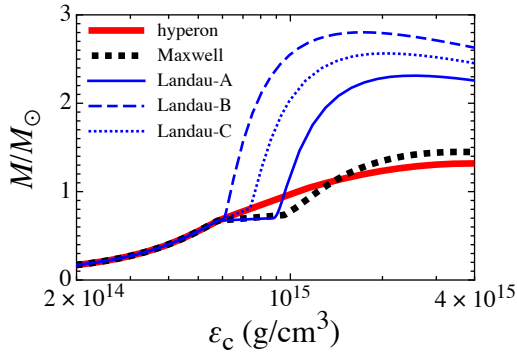} 
\end{center}
\caption{
Stellar mass as a function of the central energy density, $\varepsilon_c$, for the stellar models constructed with the different EOSs. 
}
\label{fig:rho-M}
\end{figure}

We also show the stellar masses as a function of the central energy density in Fig. \ref{fig:rho-M}, where the solid, broken, and dotted lines correspond to the stellar models constructed with the Landau-A, -B, and -C EOSs, respectively. For reference, we also add the stellar models constructed with the hyperon and Maxwell EOSs in Fig. \ref{fig:rho-M} with the thick solid and thick dotted lines. From this figure, we find that the central energy density constructing the maximum mass becomes smaller for the stellar model with the EOS connected to the hadronic matter at the lower energy density, i.e., $\varepsilon_c=1.74\times 10^{15}$ g/cm$^3$ for the Landau-B EOS, $\varepsilon_c=2.06\times 10^{15}$ g/cm$^3$ for the Landau-C EOS, and $\varepsilon_c=2.58\times 10^{15}$ g/cm$^3$ for the Landau-A EOS. Similarly, the central energy density constructing the stellar models with $M=2M_\odot$ becomes smaller for the EOS connected to the hadronic matter at the lower energy density, i.e., $\varepsilon_c=7.75\times 10^{14}$ g/cm$^3$ for the Landau-B EOS, $\varepsilon_c=9.82\times 10^{14}$ g/cm$^3$ for the Landau-C EOS, and $\varepsilon_c=1.35\times 10^{15}$ g/cm$^3$ for the Landau-A EOS. On the other hand, as shown in Fig. \ref{fig:Bc}, the critical magnetic field above which only LLL is occupied becomes smaller as decreasing the energy density. Thus, the stellar models constructed with the Landau-B EOS can be easier to be realized with weaker magnetic field, compared with the stellar models with the Landau-A EOS.

\begin{figure*}
\begin{center}
\begin{tabular}{cc}
\includegraphics[scale=0.5]{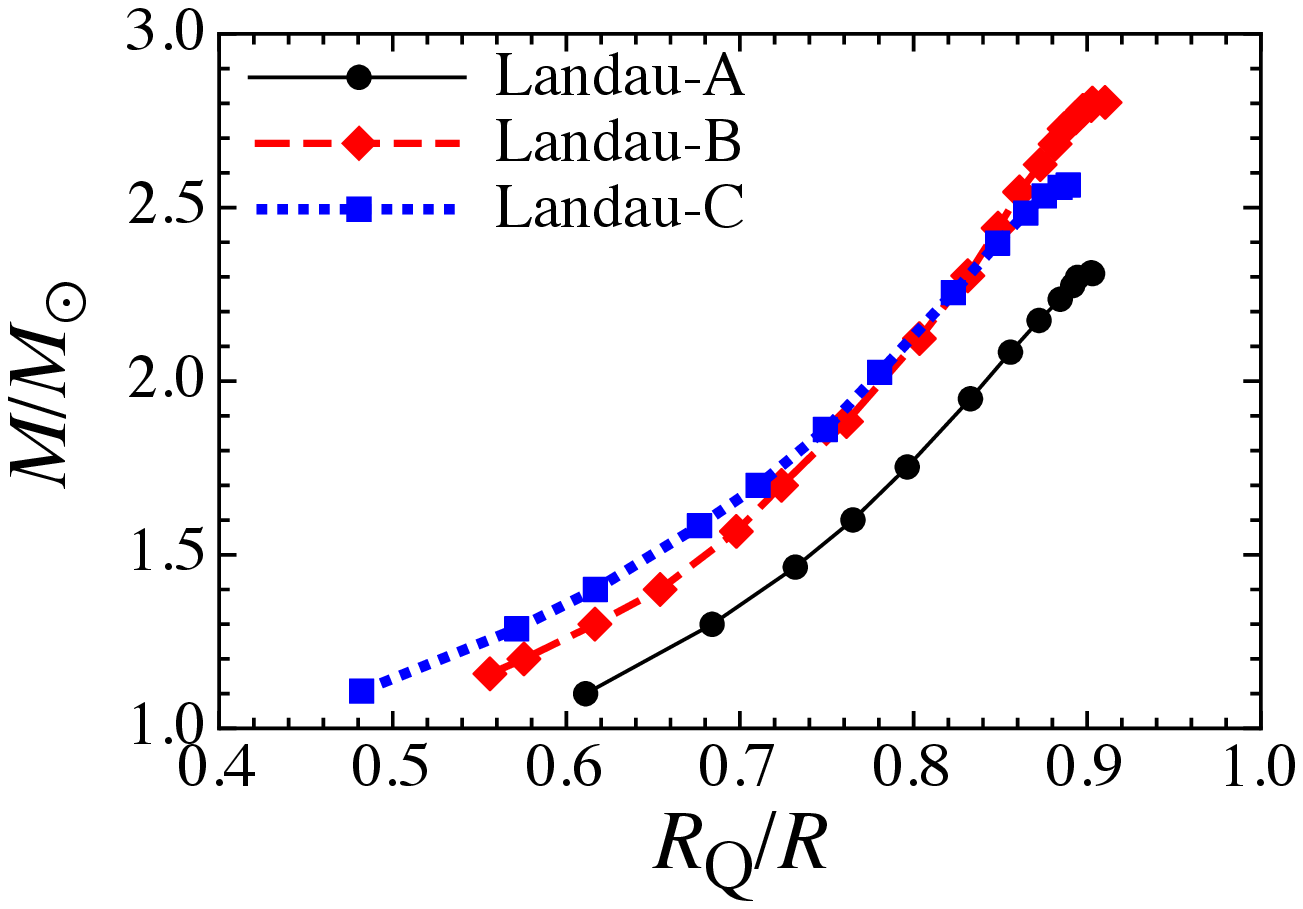} &
\includegraphics[scale=0.5]{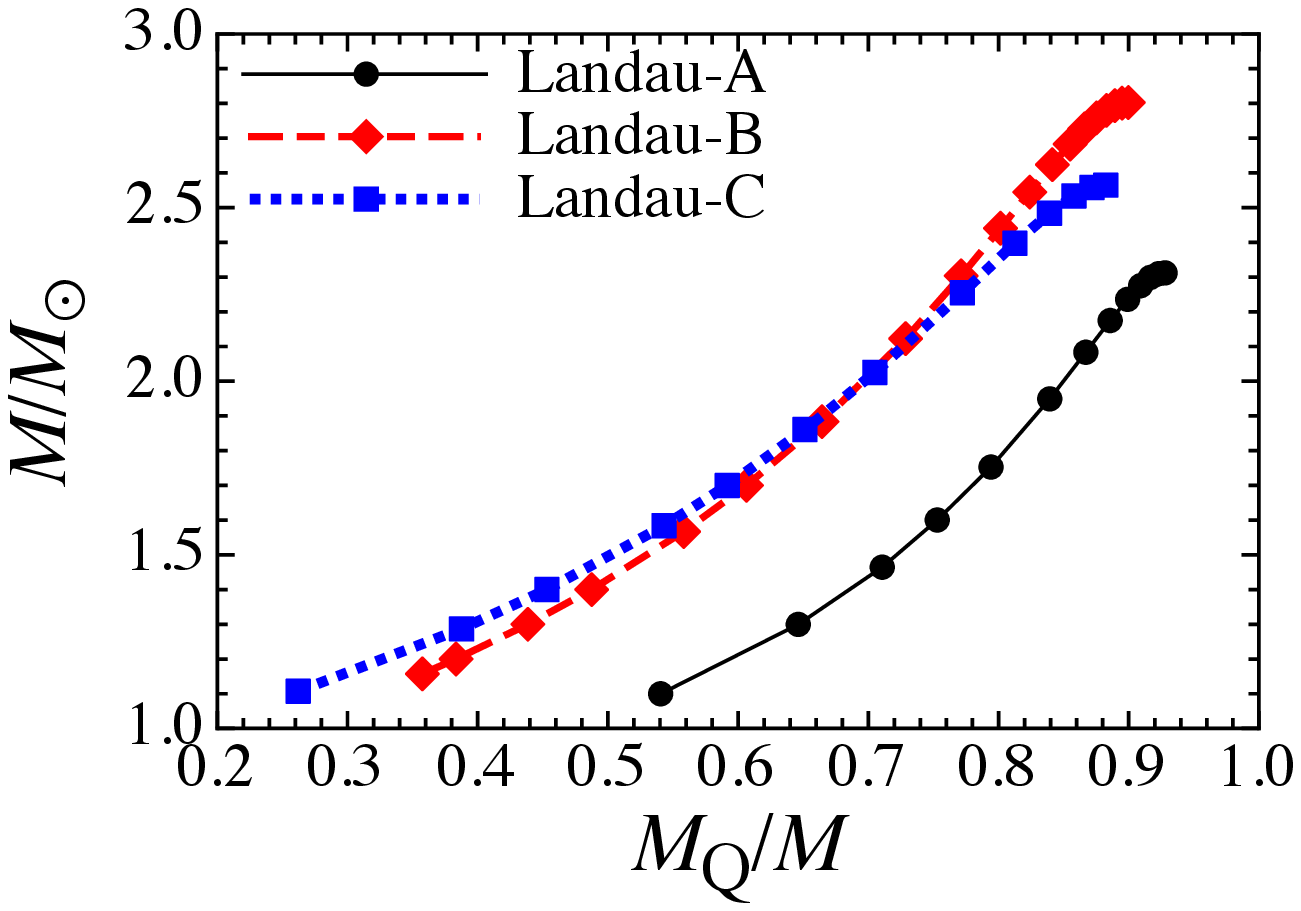}
\end{tabular}
\end{center}
\caption{
The boundary between the quark and hadronic phases in the hybrid stars constructed with the Landau-A, -B, and -C EOSs, where $R_Q$ and $M_Q$ denote the quark core radius and mass for each stellar model.  
}
\label{fig:RQ-M}
\end{figure*}

At the end, we show the boundary between the quark and hadron phases in the hybrid stars constructed with the Landau-A, -B, and -C EOSs in Fig. \ref{fig:RQ-M}, where $R_Q$ and $M_Q$ denote the quark core radius and mass. From this figure, it is obvious that, adopting the Landau EOSs, the quark phase can occupy most part of the massive hybrid stars, i.e., the quark phase for $2M_\odot$ stellar models becomes $\gsim 78\%$ of the stellar radius and $\gsim 70\%$ of the stellar mass. This is a noteworthy feature of the massive hybrid stars constructed with the Landau EOSs, because this is quite different from the massive hybrid stars suggested up to now, where the quark phase is generally quite tiny \citep{Rabhi2009,CCM2014}.

\section{Conclusion}
\label{sec:IV}

Some compact objects have a strong magnetic field, although the details of the distribution of magnetic field inside the stars are still uncertain. Maybe, one should consider the effects of such strong magnetic field on the structure of compact objects. We particularly focus on the hybrid  stars in this paper, i.e., the core region of neutron stars is composed of quark matter. We estimate that the critical magnetic field strength above which quarks only occupy LLL (and the second Landau level), which is shown that $B\sim 10^{19}$ Gauss. We also find that the equation of state (EOS) in the phase of LLL can be expressed as $P=\varepsilon-2{\cal B}$ independently of the magnetic field strength, where ${\cal B}$ denotes the bag constant. We remark that this is the limit of a stiff EOS, i.e., the sound speed becomes equal to the speed of light. We derive these results by using the MIT bag model, where quarks move almost freely, so that our findings should be relevant especially in the core region due to the asymptotic freedom of QCD, while there may be added non-perturbative effects in the moderate density region. Further consideration on the EOS of quark matter based on QCD is a future subject.

On the other hand, in general, the maximum mass of hybrid star comes to be smaller than the observed maximum mass, i.e., $2M_\odot$, because the introduction of quark matter makes the EOS soft. Even if one can construct the massive hybrid star, the quark region is generally quite tiny. However, owing to the effect of the Landau levels in the quark phase, we are successful to construct the massive hybrid quark stars occupied in large part by quark matter, which can be larger than $2M_\odot$. Furthermore, in order to examine the qualitative behavior of hybrid stellar models, we simply consider three different connections of quark matter with the hadronic matter. As a result, we find that the stellar model constructed with EOS connected to the hadronic matter at the lower energy density can realize more massive stellar model with smaller central energy density. In this paper, we consider simple stellar models as a first step, where we neglect the magnetic pressure and the deformation of stellar shape. Such additional effects will be taken into account elsewhere. In addition, one might also have to consider the hadron-quark mixed phase in the more realistic stellar models \citep{TSHT2007,Yasutake2009}. At any rate, one could see the properties of such phase modified by the strong magnetic field via the observations of the stellar oscillations \citep{Sotani2007,Sotani2008,Sotani2009}, which tells us an additional information about the strongly magnetized compact objects.

We are grateful to N. Yasutake and T. Takatsuka for their fruitful discussions, and also to our referee for reading carefully and giving valuable comments.
This work was supported in part by Grants-in-Aid for Scientific Research on Innovative Areas through 
No.\ 24105008 provided by MEXT, and by Grant-in-Aid for Young Scientists (B) through No.\ 26800133 provided by JSPS.

\appendix
\section{Stellar Models with Dipole Magnetic field}   
\label{sec:appendix_1}

In this appendix, we especially consider a dipole magnetic field with the ideal MHD approximation, as in \citet{KOK1999,Sotani2007}. That is, an axisymmetric, poloidal magnetic field produced by a four-current $J_\mu=(0,0,0,J_\phi)$ is considered, where the electromagnetic four-potential becomes very simple such as $A_\mu=(0,0,0,A_\phi)$. From the Maxwell equations $F^{\mu\nu}_{\ \ ;\nu}=4\pi J^\mu$ with the expansions of $A_\phi$ and $J_\phi$ as
\begin{eqnarray}
   A_\phi(r,\theta) &=& \sum a_\ell(r) \sin\theta\partial_\theta P_\ell(\cos\theta), \\
   J_\phi(r,\theta) &=& \sum j_\ell(r) \sin\theta\partial_\theta P_\ell(\cos\theta), 
\end{eqnarray}
one can obtain the elliptic equation describing the $\ell$-th order potential $a_\ell$;
\begin{equation}
  a_\ell'' + (\Phi' - \Lambda')a_\ell' - \frac{\ell(\ell+1)}{r^2}e^{2\Lambda} a_\ell = -4\pi e^{2\Lambda} j_\ell,
  \label{eq:da1}
\end{equation}
where a prime and $\partial_\theta$ denote the partial derivative with respect to $r$ and $\theta$, respectively, while $P_\ell(\cos\theta)$ is the $\ell$-th order Legendre polynomial. For a dipole magnetic field, i.e., $\ell=1$, the potential outside the neutron star, $a_1^{(\rm out)}$, is analytically determined by setting $j_1^{\rm (out)}=0$ and $e^{2\Phi}=e^{-2\Lambda}=1-2M/r$ \citep{WS1983};
 \begin{equation}
   a_1^{\rm (out)}(r) = -\frac{3\mu_b r^2}{8M^3}\left[\ln\left(1-\frac{2M}{r}\right) + \frac{2M}{r} + \frac{2M^2}{r^2}\right],
 \end{equation}
where $\mu_b$ is the magnetic dipole moment observed at infinity. On the other hand, the potential inside the neutron star, $a_1^{(\rm in)}$, is obtained by numerically solving Eq. (\ref{eq:da1}) in such a way that $a_1^{(\rm in)}$ should be connected to $a_1^{(\rm out)}$ at the stellar surface, where the current distribution is determined by the integration condition, i.e., $j_1^{\rm (in)}(r)=f_0 r^2(\varepsilon+P)$ \citep{KOK1999}. At last, the mantic field is determined via $a_1(r)$ as
\begin{eqnarray}
   B_r = \frac{2a_1}{r^2} e^{\Lambda}\cos\theta  \ \ {\rm and}\ \ 
   B_\theta = -a_1' e^{-\Lambda}\sin\theta,
\end{eqnarray}
which leads to the strength of magnetic field as
\begin{equation}
  B = \frac{1}{r^2}\left[4a_1^2\cos^2\theta + a_1'^2r^2e^{-2\Lambda}\sin^2\theta\right]^{1/2}.
\end{equation}
We remark that since $a_1(r)$ becomes $a_1(r)=\alpha r^2 + {\cal O}(r^4)$ in the vicinity of the stellar center, one can show that the magnetic field strength at $r=0$ is $B_0=2\alpha$.

\begin{figure}
\begin{center}
\includegraphics[scale=0.53]{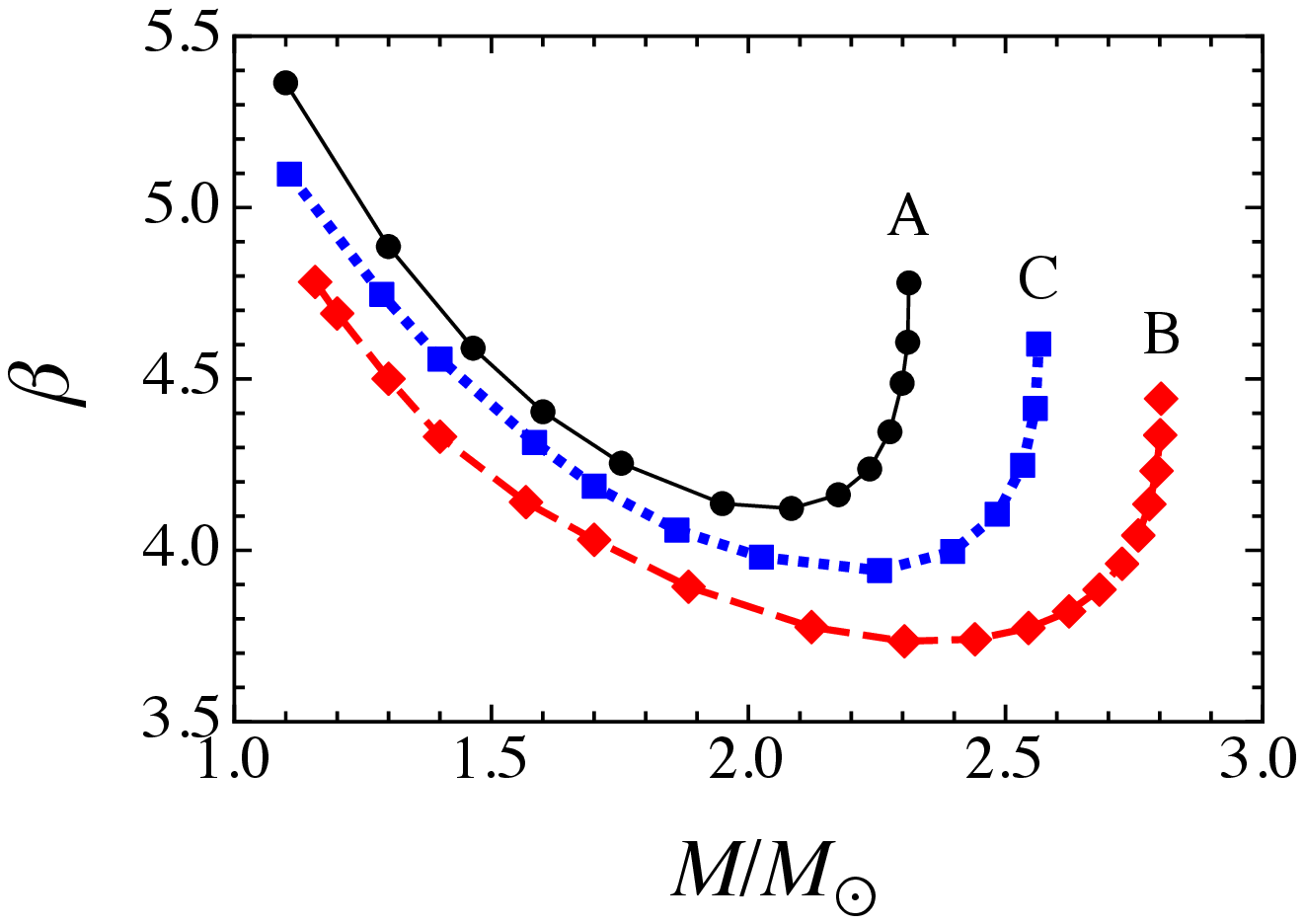} 
\end{center}
\caption{
The proportionality factor $\beta$ as a function of stellar mass constructed with the Landau-A, -B, and -C EOSs.
}
\label{fig:beta}
\end{figure}

After the calculations of the dipole magnetic distribution for the stellar model constructed with the Landau-A, -B, and -C EOSs, we find that the magnetic field strength at the stellar center is proportional to the strength at the stellar surface as
\begin{equation}
  B_0 = \beta B_p,
\end{equation}
where $B_p$ is the magnetic field strength at the stellar surface of the poles ($r=R$ and $\theta=0$), while $\beta$ is a proportionality factor. Figure \ref{fig:beta} shows the value of $\beta$ as a function of stellar mass, where the solid line with circles, broken line with diamonds, and dotted line with squares correspond to the stellar models constructed with the Landau-A, -B, and -C EOSs. From this figure, we can see that $B_0$ becomes at most five times larger than $B_p$. That is, the central strength of magnetic field can not reach to the critical magnetic field strength to realize the Landau effects in quark matter phase even for magnetar with $B_p\simeq 10^{16}$ Gauss, if the magnetic field configuration inside the star would be simply dipole.


\end{document}